\documentclass[reprint,aps,epsfig,superscriptaddress,twocolumn]{revtex4}
\usepackage{bm}
\usepackage{times}
\usepackage[dvips]{graphicx}
\usepackage{mathrsfs}
\usepackage[intlimits]{amsmath}
\usepackage{textcomp}
\usepackage[colorlinks, citecolor=red]{hyperref}

\newcommand{\ket}[1]{\vert #1 \rangle}

\begin{document}

\title{Fast delivery of heralded atom-photon quantum correlation over $12\,$km fiber through multiplexing enhancement}

\author{Sheng Zhang}
\altaffiliation{These authors contributed equally to this work}
\affiliation{Center for Quantum Information, IIIS, Tsinghua University, Beijing 100084,
PR China}

\author{Jixuan Shi}
\altaffiliation{These authors contributed equally to this work}
\affiliation{Center for Quantum Information, IIIS, Tsinghua University, Beijing 100084,
PR China}

\author{Yibo Liang}
\affiliation{Center for Quantum Information, IIIS, Tsinghua University, Beijing 100084,
PR China}

\author{Yuedong Sun}
\affiliation{Center for Quantum Information, IIIS, Tsinghua University, Beijing 100084,
PR China}

\author{Yukai Wu}
\affiliation{Center for Quantum Information, IIIS, Tsinghua University, Beijing 100084,
PR China}
\affiliation{Hefei National Laboratory, Hefei 230088, PR China}

\author{Luming Duan}
\email{lmduan@tsinghua.edu.cn}
\affiliation{Center for Quantum Information, IIIS, Tsinghua University, Beijing 100084,
PR China}
\affiliation{Hefei National Laboratory, Hefei 230088, PR China}

\author{Yunfei Pu}
\email{puyf@tsinghua.edu.cn}
\affiliation{Center for Quantum Information, IIIS, Tsinghua University, Beijing 100084,
PR China}
\affiliation{Hefei National Laboratory, Hefei 230088, PR China}

\begin{abstract}
Distributing quantum entanglement between distant parties is a significant but difficult task in quantum information science, as it can enable numerous applications but suffers from exponential decay in the quantum channel. Quantum repeater is one of the most promising approaches towards this goal. In a quantum repeater protocol, it is essential that the entanglement generation speed within each elementary link is faster than the memory decoherence rate, to enable the scale-up of the quantum repeater by connecting neighboring repeater segments. This stringent requirement has not been implemented over a fiber of metropolitan scale so far. As a step towards this challenging goal, in this work we experimentally realize multiplexing-enhanced generation of heralded atom-photon quantum correlation over a $12\,$km fiber. We excite the memory modes in a multiplexed quantum memory successively to generate $280$ pairs of atom-photon quantum correlations with a train of photonic time-bin pulses filling the long fiber. After successful detection of a heralding signal, the excited memory mode can be identified and retrieved into idler photons on demand with either fixed or variable storage time. With the multiplexing enhancement, the heralding rate of atom-photon correlation can reach $1.95\,$kHz, and the ratio between the quantum correlation generation rate to memory decoherence rate can be improved to $0.46$ for a fiber length of $12\,$km, which is so far the best for long fiber length ($>$$10\,$km) to our knowledge. This work therefore constitutes an important step towards the realization of a large-scale quantum repeater network.
\end{abstract}

\maketitle


Quantum repeater is a most promising way to distribute quantum entanglement between two distant locations~\cite{BDCZ, DLCZ, Gisin}, which can then be used for various applications such as quantum communication~\cite{E91, oxford qkd, weinfurter qkd}, networked quantum sensing~\cite{ye and lukin, repeater telescope} and distributed quantum computing~\cite{distributed}. Recently, many achievements have been realized in this direction. Heralded entanglement generation in an elementary link of quantum repeater has been realized at both short and long lengths~\cite{weinfurter qkd, oxford ion, monroe, rempe, hansondelivery, pan2008,2021nature, queqiao, sc, bao3nodes, 33km, nvlong, 230m, lukin reflection, weinfurter1km,hanson3node}. Memory enhanced connection of two atom-photon entanglements~\cite{lanyon connection, lattice} and $3$-node quantum network~\cite{bao3nodes,hanson3node} have also been implemented recently. Besides, there are also many achievements following the approach of memory-less quantum repeater~\cite{photonic_repeater, pan19repeater, photonicexperiment}.

A critical requirement for the scaling of a quantum repeater is that the expected time cost of entanglement generation in each elementary link should be shorter than the memory coherence time~\cite{monroe, hansondelivery}, which sets a stringent criterion to the performance of a quantum memory in both coherence time and the ability to interface atom-photon entanglement or correlation with high efficiency. A figure of merit to quantitatively characterize the ability of delivering quantum entanglement within memory coherence time is the quantum link efficiency $\eta_\text{\,link}$ defined in~\cite{hansondelivery}, which can be decomposed as:
\begin{equation}
\eta_\text{\,link}=\frac{T_\text{coh}}{T_\text{ent}}=\frac{T_\text{coh}}{\frac{2L}{cNp}}=\frac{c}{2L}T_\text{coh}pN
\end{equation}
where $T_\text{coh}$ is the memory coherence time, $T_\text{ent}$ is the expected time cost of generating an atom-atom entanglement in an elementary link, $L$ is the distance between the memory and the center detection station (half length of the elementary link, see Fig.~1a~and~1b), $c$ is the speed of light in fiber, $N$ is the effective enhancement through multiplexing, and $p$ is the success probability of each trial. It is argued that the threshold for deterministic delivery of quantum entanglement is $\eta_\text{\,link}\gtrsim0.83$~\cite{hansondelivery}. It is also noteworthy that $\eta_\text{\,link}$ is twice the expected number of heralded atom-photon correlation generated within memory coherence time over $L$, if Type I protocol (single photon interference) is used~\cite{DLCZ,2021nature, bao3nodes, Gisin, type I, hansondelivery, hanson3node} (by receiving clicks from both detectors at the detection station, see Fig.~1a).  As illustrated in Eq.~(1), it is crucial to improve $N$, $p$, and $T_\text{coh}$ to achieve a high $\eta_\text{\,link}$.

\begin{figure*}
  \centering
  \includegraphics[width=18cm]{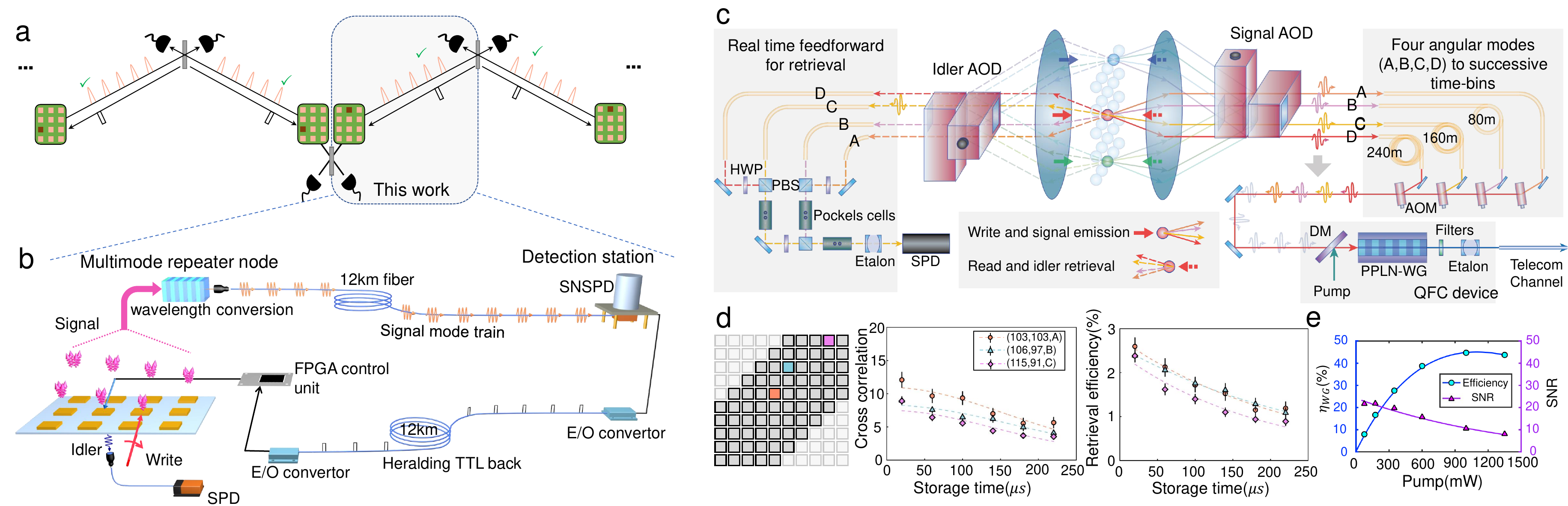}\\
  \caption{ Experimental setup and scheme.
  \textbf{a}, The position of this work in a multiplexed quantum repeater protocol. This experiment represents half of the heralded entanglement generation in a long elementary link of a multiplexed quantum repeater. \textbf{b}, The protocol of this experiment. We use a full protocol that the heralding signal arrives $\frac{2L}{c}$$=$$120\,\mu$s after the excitation to simulate the real-world application. In this protocol, pairs of quantum correlations between signal photon modes and corresponding memory (spin-wave) modes are generated successively. The signal modes are converted to C band and sent into a $12\,$km fiber one by one in time-bin pulses and detected after fiber transmission. The successful detection of signal photon is converted to TTL pulse at the detector and is further sent back to the memory through another $12\,$km fiber with the help of two E/O (electric-optical) converters. The memory receives the heralding TTL pulse and reads out the corresponding memory mode according to the arrival time of the TTL pulse. The detection station and the memory is $5\,$m apart in a lab. The fibers are also in the same lab. \textbf{c}, The detailed experimental setup. We combine $70$ spatial and $4$ angular dimensions into totally $280$ time-bin modes. Here each combination of the spatial and angular modes can be individually addressed by the 2D AOD system with the help of AOMs (acoustic-optical modulators) and EOMs (electrical-optical modulators). The signal mode at $795\,$nm is converted to $1546\,$nm on a PPLN waveguide. \textbf{d}, The performance of the memory array. We choose $3$ memory modes with each locating at different representative regions of the memory array and emitting at different emission angles, and demonstrate corresponding retrieval efficiency (including all losses) and cross correlation $g_{s,i}=\frac{p_{s,i}}{p_sp_i}$ in each mode with varying storage time. The corresponding storage time of the three modes are $227\pm21\,\mu$s, $261\pm22\,\mu$s and $190\pm22\,\mu$s, respectively. \textbf{e}, The performance of wavelength conversion. The device efficiency (circle) of the PPLN waveguide and corresponding signal-to-noise ratio (triangle) are demonstrated at a success probability of $\bar{p}=0.04\%$. The highest device efficiency is about $45\%$ with a pumping laser about $1\,$W, and the end-to-end efficiency is about $12\%$ due to multiple stages of filtering and fiber connections.
   }
\end{figure*}

 This requirement has already been fulfilled on a laboratory scale ($\sim$$10\,$m)~\cite{monroe, hansondelivery, hanson3node}, but is yet to be fulfilled if the elementary link reaches metropolitan size ($L>10\,$km), due to significantly lower repetition rate and success probability for longer distance. As the flying qubit and heralding signal needs to experience a round-trip travel of $2L/c$ ($>$$100\,\mu$s, see Fig.~1a~and~1b) to herald the success of the entanglement generation in each excitation trial, the repetition rate is limited to $<$$10\,$kHz in the entanglement generation stage. To achieve a faster repetition thus higher entanglement generation rate, multiplexed quantum repeater is proposed and implemented~\cite{07prl, Gisin, simon,2021nature,queqiao, wanghai, lanyon multiplexing, riedmattenlongfiber}. A significant enhancement of $62$ modes in a $50\,$m fiber has been demonstrated with rare earth ion ensemble recently~\cite{2021nature}, while the coherence time ($\sim$$25\,\mu$s) needs to be improved for longer fiber length. Another approach is to improve the success probability $p$ in each trial, and efforts on cavity enhancement~\cite{230m,lanyon connection, lanyon multiplexing, keller, rempe, lukin reflection} and Rydberg blockade~\cite{bao rydberg} have been demonstrated recently towards this goal. The cavity-enhanced trapped ion system can generate about $0.35$ expected atom-photon entanglement within memory coherence over $25\,$km fiber~\cite{lanyon connection}, which, to our knowledge, is the current record for the rate to generate heralded atom-photon entanglement or correlation within memory coherence time. Meanwhile, issues like unwanted spontaneous emission~\cite{230m,keller} and cavity jitter limit the indistinguishability of the heralding photon, which need to be suppressed in the future to guarantee high-quality ion-ion entanglement without harming the efficiency significantly~\cite{230m}.  So far, the quantum link efficiency $\eta_\text{\,link}$ is still below $0.01$ for heralded entanglement generation at a fiber length over $1\,$km~\cite{bao3nodes, lukin reflection, 33km, nvlong, weinfurter1km}, to the best of our knowledge.

In this work, as a step towards this outstanding goal, we achieve fast delivery of heralded atom-photon quantum correlation with a fiber length of $12\,$km via a multiplexing-enhanced quantum repeater protocol. The generation rate is enhanced by $280$ time-bin modes and can reach $0.46$ expected remote atom-photon quantum correlation within memory coherence time, which is so far the best to our knowledge. This corresponds to a $\eta_\text{\,link}$$\,$$=$$\,$$0.92$ if two such heralded atom-photon quantum correlations are employed to generate a $24\,$km atom-atom entanglement via single photon interference~\cite{DLCZ,2021nature, bao3nodes, Gisin, type I, hansondelivery, hanson3node} in the future, which will be close to the scale-up requirement to build a multi-node quantum repeater.  The distribution rate of remote quantum correlation equals to $1.95\,$kHz during each $240\,\mu$s protocol, or an averaged rate of $187\,$Hz if all time costs are counted, which are also the records for long fiber length ($>$$10\,$km) in heralded style. Furthermore, the excited memory modes can be retrieved into idler photons with either fixed or variable storage time on demand, which is suitable for applications such as the synchronization and connection of two asynchronously entangled quantum repeater segments. It is also noteworthy that the built-in random-access addressing ability in our memory can support nearly arbitrary quantum network protocols~\cite{multipurpose, 225}. Finally, the signal modes are detected after transmission in the long fiber and the classical heralding TTLs also travel in another long fiber to reach the memory, which closely resemble the real-world protocol in a long elementary repeater link. With all these achievements, this work constitutes an important step towards a long-distance quantum repeater.

\section{Experimental protocol and results}

\begin{figure}
  \centering
  \includegraphics[width=8.5cm]{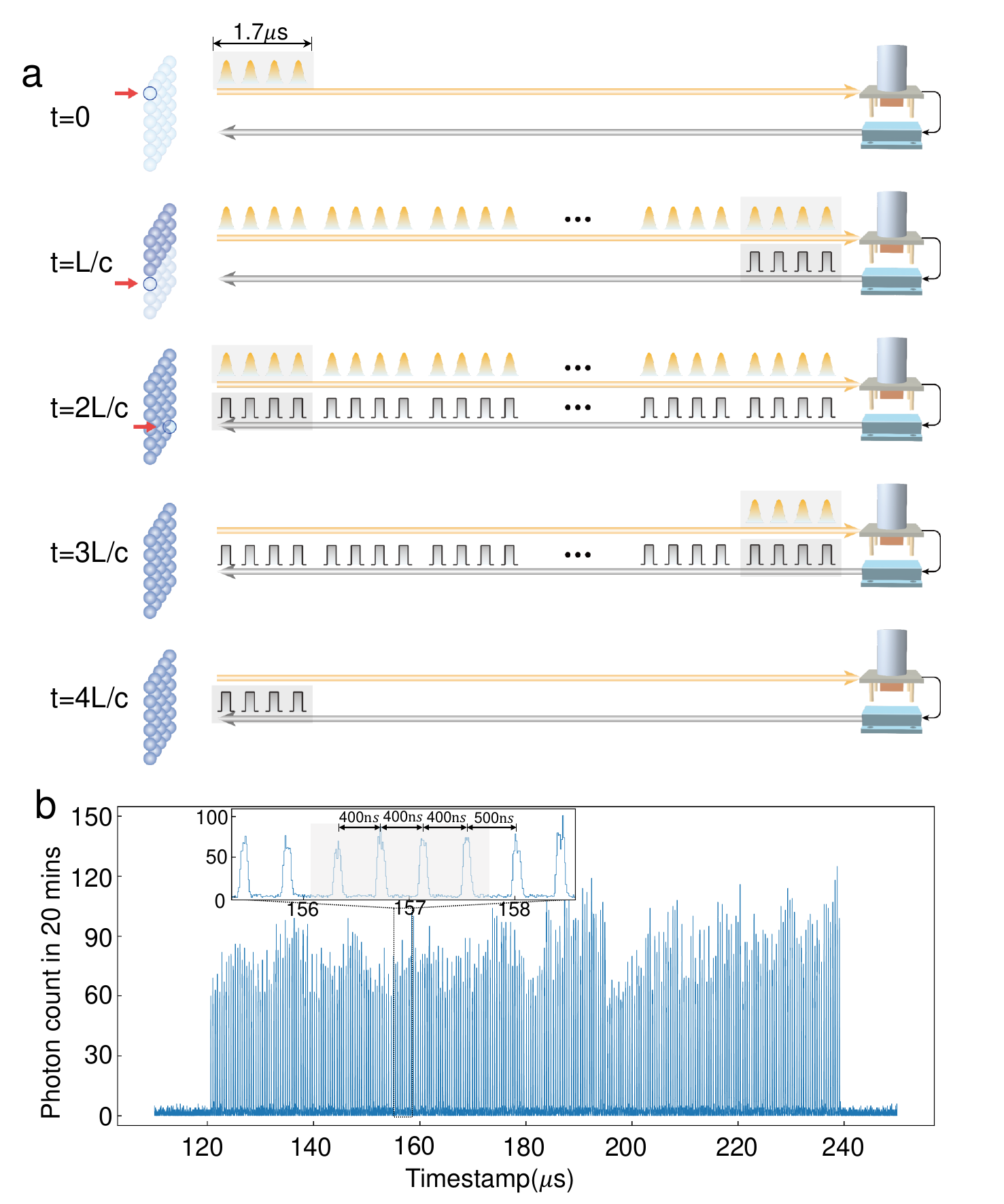}\\
  \caption{ Remote heralding with $280$ time-bin modes.
  \textbf{a}, The heralding protocol of remote atom-photon quantum correlation over an $L=12\,$km fiber enhanced by $280$ time-bin modes. The first $4$ modes are generated at $t=0$ when the addressed write beam excites the first memory cell. The time-bin modes of signal photon are travelling in the long fiber as a long pulse train, with an interval at $400\,$ns (between modes from the same cell) or $500\,$ns (between the $4$th mode from previous cell and $1$st mode from current cell). \textbf{b}, The histogram of the arrival time for the heralding TTL received by the memory, at an average success probability of $\bar{p}=0.18\%$. Note that the excitation probability here is higher than in Fig.~1e, thus higher signal-to-noise ratio is achieved. The inset is a zoom-in in the long time-bin pulse train, the shaded $4$ modes are from the same memory cell. The temporal width of each signal mode is about $100\,$ns.
   }
\end{figure}

\begin{figure}
  \centering
  \includegraphics[width=8.7cm]{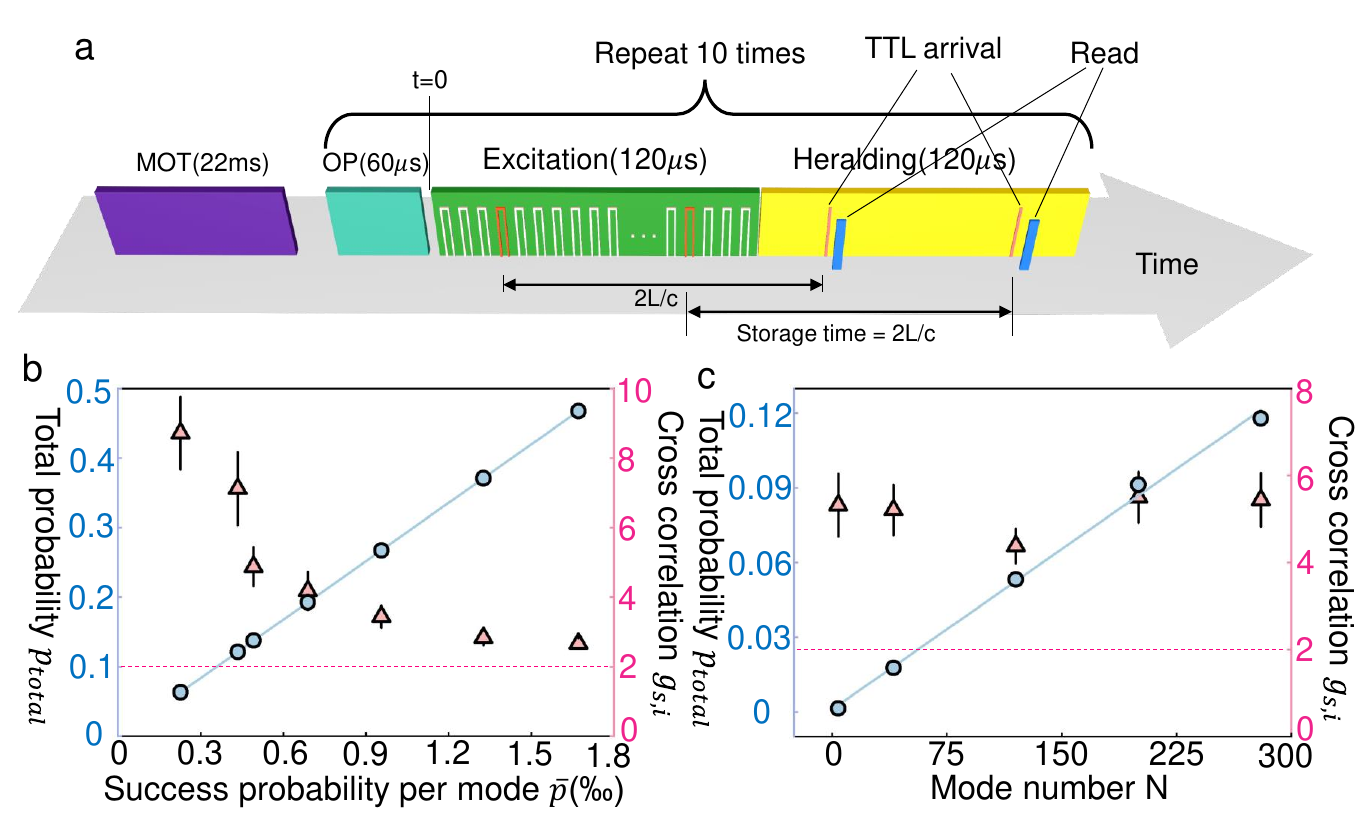}\\
  \caption{ The protocol and results when the storage time is fixed.
  \textbf{a}, The protocol with the storage time fixed during the retrieval. $10$ rounds of repetitions lasting about $3\,$ms follows the $22\,$ms MOT loading stage. In each round, the generation of heralded quantum correlation starts after the $60\,\mu$s optical pumping. During the excitation stage, we start to excite the first memory cell at $t=0$, and it costs about $120\,\mu$s to excite all the $280$ modes. The heralding stage lasts from $t=120\,\mu$s to $t=240\,\mu$s. Once a heralding TTL pulse is received, the memory immediately read the corresponding spin-wave mode out. In this case, the storage time is always fixed to $130\,\mu$s. The memory can register and retrieve multiple heralded quantum correlations (at most $3$) in each round. \textbf{b}, \textbf{c}, The total probability $p_\text{\,total}$ (blue circle) and cross correlation $g_{s,i}=\frac{p_{s,i}}{p_sp_i}$ (pink triangle) with varying average success probability $\bar{p}$ and mode number $N$. The dashed line represents the upper bound of classical correlation at $2$. Here $N=280$ in b and $\bar{p}=0.42$\textperthousand\, in c.
  }
\end{figure}

We implement the multiplexed DLCZ (Duan-Lukin-Cirac-Zoller) quantum memory by combining spatial~\cite{multipurpose, 225, lan} and angular~\cite{wavevector, wanghai} dimensions. We first realize a $10\times10$ spatially multiplexed memory array via $2$D AOD addressing units~\cite{multipurpose, 225}, as shown in Fig.~1c. Here each memory cell is a micro-ensemble of Rb$^{87}$ atoms, and all the cells are prepared via an optical pumping first, which pumps the atoms to $|5S_{1/2},F=1, m_F=+1\rangle$, followed by a microwave $\pi$-pulse to transfer the population to $|5S_{1/2},F=2, m_F=0\rangle$ (see supplemental material). We excite $70$ memory cells one by one via the individually addressed write beam with a switching time of $1.7\,\mu$s (see Fig.~1b), which drives spontaneous Raman transition to create the quantum correlation between signal photon at $795\,$nm and collective atomic excitation (spin-wave) based on two clock states $|g\rangle\equiv|5S_{1/2},F=2, m_F=0\rangle$ and $|s\rangle\equiv|5S_{1/2},F=1, m_F=0\rangle$ following the DLCZ protocol~\cite{DLCZ, lattice}. Here four emission angles of signal photon are collected in a single write process, and quantum correlations are generated in these four pairs of signal modes and corresponding spin-wave modes~\cite{wanghai, wavevector} during the excitation of each memory cell. Each of the four angular signal modes has an angle of $0.3^{\circ}$ to the write beam, and the averaged coherence time for each spin-wave mode is $235\,\mu$s.

After the excitation, the $4$ angular modes of signal photon go through the signal AOD simultaneously but are separated spatially (see Fig.~1c and supplemental material). We further pick out each signal mode after AOD and send them into delay fibers with different lengths, then combine them into $4$ successive time-bin modes in a single fiber with a time interval of $400\,$ns. In this way, we can generate quantum correlations between $280$ time-bin signal modes and corresponding spin-wave modes by successively excite the $70$ memory cells in the array, with the $280$ signal modes enter the same fiber one by one. The locations of the $70$ cells in the $10\times10$ array are shown in Fig.~1d, and the reason for this selection is explained in the supplemental material. The time-bin signal modes at $795\,$nm further go through a PPLN (Periodically poled lithium niobate) waveguide and are converted to $1546\,$nm via DFG (Difference frequency generation)~\cite{Ikuta,lanyon connection, bao3nodes,33km, lukin reflection, eschner, riedmattenconversion, quraishi}. The conversion efficiency and signal-to-noise ratio (SNR) are illustrated in Fig.~1e. Note that the SNR will increase with higher excitation probability of signal photon. The time-bin train of converted signal modes are further sent through the $12\,$km fiber to the detection station ($\sim$$5\,$m away from the memory), where the signal modes are converted to TTL (Transistor-Transistor Logic) pulses at the SNSPD (superconducting nanowire single-photon detector).

To notify the memory, the heralding TTL  pulses are converted to optical pulses by an E/O converter and sent back to the memory via another $12\,$km fiber, as shown in Fig.~1b and 2a. After the optical pulses arrives at the memory, they are converted back to TTL pulses via another E/O converter. The FPGA control unit gets the heralding signal and identifies which spin-wave mode (both spatial and angular) is excited depending on the TTL arrival time, and reads out the corresponding spin-wave mode to an idler photon, which can be used for verification of the distant quantum correlation or further connection of neighboring repeater segments. This protocol demonstrated here simulates a real-world protocol for heralded generation of distant atom-photon quantum correlation, which can be further employed to generate atom-atom entanglement in an elementary repeater link via single photon interference~\cite{DLCZ,2021nature, bao3nodes, Gisin, type I}.

\begin{figure}
  \centering
  \includegraphics[width=8.7cm]{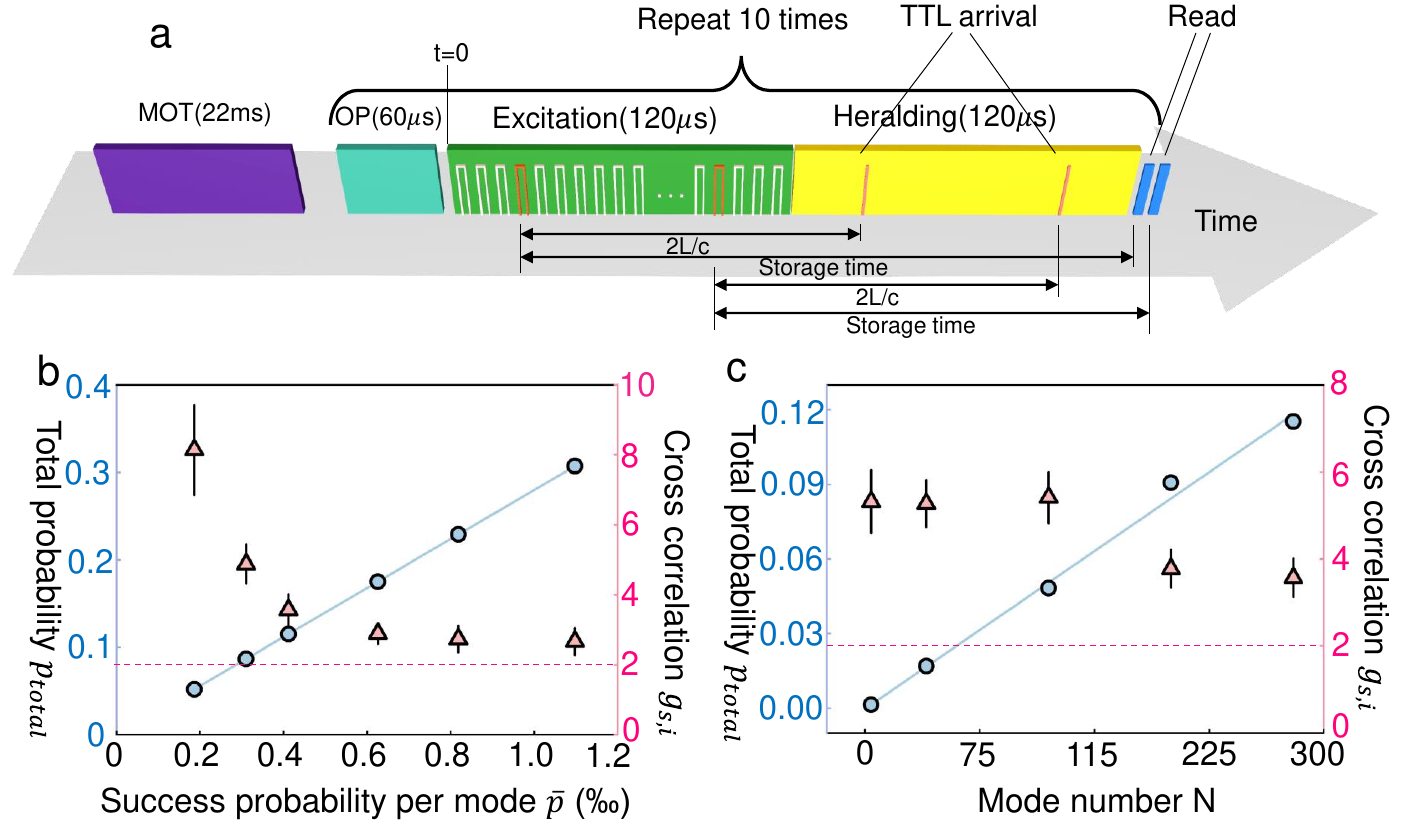}\\
  \caption{ The protocol and results when the retrieval time is user-defined.
  \textbf{a}, The protocol where the retrieval time is fixed to the end of heralding stage. In the heralding stage, once a heralding TTL pulse is received, the memory registers this event, wait and read out the corresponding spin-wave mode at a pre-defined time ($250\,\mu$s here). In this case, the storage time is random over the range from $130\,\mu$s to $250\,\mu$s. The protocol can also register and retrieve multiple heralded quantum correlations (at most $3$). \textbf{b}, \textbf{c}, The total probability $p_\text{\,total}$ and cross correlation $g_{s,i}$ with varying average success probability $\bar{p}$ and mode number $N$. The dashed line represents the upper bound of classical correlation at $2$. Here $N=280$ in b and $\bar{p}=0.43$\textperthousand\,  in c.}
\end{figure}

When $280$ signal modes are used, we need $70$$\times$$1.7$$=$$119\,\mu$s to finish the excitation process, which approximately corresponds to $\frac{2L}{c}=120\,\mu$s, the round-trip travel time in both of the $12\,$km fibers for qubit and TTL. This means the heralding TTL pulse for the first signal mode arrives at the memory right after the last memory cell is excited, as illustrated in Fig.~2a. The $12\,$km qubit fiber is fully filled from $t=\frac{L}{c}$ to $\frac{2L}{c}$. The control unit starts receiving the heralding TTL since $t=\frac{2L}{c}$, and this process keeps going until $t=\frac{4L}{c}=240\,\mu$s. The histogram for the arrival time of heralding TTLs is registered and illustrated in Fig.~2b, which also represents the success probability of each signal modes. Here we can execute $280$ trials in each $240\,\mu$s round, with an effective repetition rate at $1.17\,$MHz, which is more than $100$ times faster than the maximum repetition rate $\frac{c}{2L}=8.3\,$kHz if there's no multiplexing enhancement.

After the successful heralding in any of the $280$ modes, we read out the stored spin-wave mode for further applications. Here we demonstrate the retrieval of the quantum information stored in the quantum memory in two different styles, which are (\textbf{i}) reading out the corresponding memory mode immediately after receiving a heralding TTL, and (\textbf{ii}) reading out the corresponding memory mode at a predefined time no matter when the heralding TTL is received. It is also noteworthy that our memory can perform an arbitrary protocol in the reading as it can be performed in a random-access way~\cite{225, multipurpose}, not limited to these two styles demonstrated here.

In the first read-out style, the storage time is fixed, but the read-out time is random. As illustrated in Fig.~3a, the control unit will identify and read out the excited spin-wave mode immediately after receiving a heralding TTL in the $120\,\mu$s heralding stage. This is achieved by dynamically controlling the RF (radio frequency) signal sent to the AOD depending on the TTL arrival time, which can be completed in  $10\,\mu$s after receiving the heralding TTL (see supplemental material). Thus the storage time is always $\frac{2L}{c}$$+$$10$$=$$130\,\mu$s for each of the $280$ modes. If more than one TTLs are received in the heralding stage, we can either read all of them out or immediately terminate the heralding stage right after the first heralding TTL is received. Here we choose to read out at most $3$ excited spin-wave modes during each heralding stage for accelerated data collection. Idler photon from different spin-wave modes are also combined into a single fiber via an EOM (electro-optical modulator) network, as shown in Fig.~1c.

Here we use $p_\text{\,total}=\sum_{i=1}^Np_i$ to characterize the total success probability in one round of excitation, where $p_i$ is the success probability of detecting the $i$th signal mode at the detection station (and $i$th heralding TTL at the memory), and $N$ is the number of total used modes.  We also use the cross-correlation $g_{s,i}=\frac{p_{s,i}}{p_sp_i}$ to characterize the quality of the heralded atom-photon quantum correlation (see supplemental material for details). In the DLCZ quantum repeater protocol, the efficiency and the quality are in a trade-off~\cite{DLCZ}, and here we demonstrate the total efficiency $p_\text{\,total}$ and cross-correlation $g_{s,i}$ with a varying average success probability $\bar{p}=\frac{1}{N}p_\text{\,total}$ in Fig.~3b. As shown in Fig.~3b, the $p_\text{\,total}$ increases linearly with $\bar{p}$ and $g_{s,i}$ decays with $\bar{p}$. When $\bar{p}=0.167\%$, we can achieve a total success probability as high as $p_\text{\,total}=0.47$, with the cross correlation at $g_{s,i}=2.67\pm0.15$, which provides clear evidence for quantum correlation as the classical bound is $2$~\cite{kimble2006}. The whole time cost of this protocol is the sum of the excitation stage $120\,\mu$s and the heralding stage $120\,\mu$s, which equals to $240\,\mu$s. Here the $60\,\mu$s optical pumping can be combined into the overhead like MOT loading because optical pumping is only needed once before the beginning of the $240\,\mu$s protocol, and no longer needed in the middle of the protocol (from $t$$=$$0$ to $t$$=$$240\,\mu$s). This also means that we can generate heralded atom-photon correlation at a rate of $\frac{0.47}{240\,\mu\rm{s}}=1.95\,$kHz in each round of $240\,\mu$s. If we take everything (including MOT loading and optical pumping) into account, the average remote atom-photon correlation generation rate is about $187\,$Hz, with a duty cycle of $9.6\%$. Given the average coherence time $T_\text{coh}=235\,\mu$s, here we can generate $1.95\,\text{kHz}\times235\,\mu\text{s}=0.46$ expected atom-photon correlation within memory coherence time, which also corresponds to an equivalent quantum link efficiency $\eta_\text{\,link}=0.46\times 2=0.92$, if two such setups are combined to generate heralded atom-atom entanglement with single photon interference in the future.  We can also vary the number of modes used in this protocol, and the results with $\bar{p}=0.042\%$ are shown in Fig.~3c. As shown in Fig.~3c, the total success probability increases linearly with $N$, which clearly illustrates the multiplexing enhancement. When $N<280$, there will be a gap between the excitation and heralding stage, but the storage time is still fixed to $130\,\mu$s. The cells used in different $N$ are described in detail in the supplemental material.

In the second read-out style, no matter when the heralding TTL is received, the memory always reads out the stored spin-wave mode at a user-defined timestamp, and here we set this timestamp at the end of the heralding stage to cover all the $280$ modes, as shown in Fig.~4a. Note that in this protocol, the read-out time is fixed no matter which mode is excited, meanwhile the storage time is variable (for example, the storage time is $250\,\mu$s for the $1$st mode, and $130\,\mu$s for the $280$th mode). This protocol is more favorable than the first protocol for the connection between different repeater segments as it simulates the synchronization and connection between two asynchronously entangled elementary links, which requires the the quantum memory can handle a variable storage time and can be retrieved on demand. Here we can also register at most $3$ heralding TTLs and read corresponding spin-wave modes out for faster data collection. We also demonstrate the total heralding probability $p_\text{\,total}$ and cross correlation $g_{s,i}$ with varying average probability $\bar{p}$ and mode number $N$ in Fig.~4b and 4c. We can achieve $p_\text{\,total}=0.30$ with the quality of quantum correlation still above the classical bound, which corresponds to an equivalent quantum link efficiency $\eta_\text{\,link}=0.60$ in this case. The slight decay of corresponding $p_\text{\,total}$ and cross-correlation compared to fixed storage time case mainly origins from the longer storage time in this case.

\section{conclusion}

In conclusion, we experimentally realize heralded generation of atom-photon quantum correlation over $12\,$km fiber, with an multiplexing-enhanced efficiency via the use of $280$ modes. We demonstrate an expected delivery of $0.46$ atom-photon quantum correlation within memory coherence time, which represents the current state-of-the-art. Two different read-out styles are also demonstrated in this work and the quantum correlation can be generated with a record-high rate over a long fiber. In the future, we can use deployed fibers and distant quantum repeater nodes to establish quantum repeater link between spatially separated locations~\cite{nvlong, 33km, bao3nodes, 230m, riedmattenlongfiber, lukin reflection, weinfurter1km}. There is still room to improve the efficiency of wavelength conversion ($12\%$) considering the state-of-the-art is $57\%$~\cite{33km}, and this can further improve corresponding $\eta_\text{\,link}$ by several times.  We can also improve the performance of the memory by loading the memory cells into a two-dimensional optical lattice array. With the improved storage time around $50\,$ms~\cite{lattice}, the memory can achieve an equivalent quantum link efficiency over $100$, which will constitute a promising platform for building a multi-layer quantum repeater network extended over a metropolitan or continental scale in the future.

\noindent\textbf{Data availability} The data that support the findings of this study are available from the
corresponding authors upon request.

\noindent\textbf{Acknowledgements} We acknowledge helpful discussions with Xiaohui Bao, Yong Yu, Jun Li, Wei Zhang, and Zhiyuan Zhou. This work is supported by Innovation Program for Quantum Science and Technology (No.2021ZD0301102), the Tsinghua University Initiative Scientific Research Program, the Ministry of Education of China through its fund to the IIIS, and National Key Research and Development Program of China (2020YFA0309500). Y.P. acknowledges support from the start-up fund and the Dushi Program from Tsinghua University. Y.W. acknowledges support from the start-up fund from Tsinghua University.

\noindent\textbf{Author Contributions} S.Z, J.S., Y.L., Y.S. and Y.P. carried out the experiment. L.D. and Y.P. supervised the project. All the authors contributed to the discussion and the writing.

\noindent\textbf{Author Information} The authors declare no competing financial interests. Correspondence and requests for materials should be addressed to Y.P. (puyf@tsinghua.edu.cn) or L.D. (lmduan@tsinghua.edu.cn).

\onecolumngrid


\setcounter{equation}{0}
\setcounter{figure}{0}
\setcounter{table}{0}
\setcounter{page}{1}
\setcounter{section}{0}
\makeatletter
\renewcommand{\theequation}{S\arabic{equation}}
\renewcommand{\thefigure}{S\arabic{figure}}
\renewcommand{\thetable}{S\arabic{table}}

\pagebreak
\begin{center}
\Large Supplemental Material for\\\textbf{Fast delivery of heralded atom-photon quantum correlation over $12\,$km fiber through multiplexing enhancement}
\end{center}

\section{ Experimental setup and method}
In this section, we describe the experimental implementation of a spatially-multiplexed quantum memory array combined with angular degree of freedom. The experimental realization of a 2D quantum memory array with hundreds of individually addressable cells has been elaborated in our previous works~\cite{225,multipurpose}. In this work, we integrate the angular degree of freedom into the spatially-multiplexed quantum memory array to enlarge the multimode capacity. When a memory cell is excited by the focused write beam, we collect signal photons from four symmetrical angles (labeled as A,B,C,D), as depicted in Fig. S1a. The light emitted from the same cell at these four angles corresponds to distinct regions on the AOD crystal. We employ D-type mirrors to further seperate these four spatial patterns and direct them into four single-mode fibers with varying lengths, respectively. Subsequently, after passing through four fibers and four AOM switches, the four angular modes from one memory cell are transformed into four time-bin modes within the same fiber. Upon sequentially exciting all the memory cells in the 2D array $(10\times 10)$, the $100$ spatial modes and four angular modes are combined into $400$ photonic time-bin modes, suitable for implementing multiplexing enhancement over a long-distance fiber.

In addition, we need each mode to have a sufficient lifetime to store the atom-photon correlation during the long-distance fiber transmission. The coherence time of each memory cell over $500\,\mu$s has been demonstrated in our prior work. Here we follow the similar methods to extend the coherence time, employing a collinear configuration and optical pumping to the clock state. However, due to utilizing four angular modes simultaneously, achieving complete collinearity between the write beam and signal photon is impossible. In this setup, the angle between the write beam and the signal photon is $0.3^{\circ}$. The initial state $\ket{g}\equiv\ket{5S_{1/2},F=2,m=0}$ is prepared after a $60\,\mu$s optical pumping process with a bias magnetic field of $0.5\,$G. As illustrated in Fig. S2a, the whole optical pumping process consists of three parts. In the first $20\,\mu$s, we optically pump all the atoms to $\ket{5S_{1/2},F=1,m=+1}$ state using three pump lights with different frequencies and polarizations, including one $\sigma_+$ light coupling the $\ket{5S_{1/2},F=1}\rightarrow\ket{5P_{1/2},F=1}$ transition, one $\pi$ light coupling the $\ket{5S_{1/2},F=2}\rightarrow\ket{5P_{1/2},F=2}$ transition, and another $\sigma_+$ light coupling the $\ket{5S_{1/2},F=2}\rightarrow\ket{5P_{3/2},F=2}$ transition. In the second $20\,\mu$s, we execute a $\pi$ flip between the $\ket{5S_{1/2},F=1,m=+1}$ state and the $\ket{5S_{1/2},F=2,m=0}$ state using a microwave transition at a specific frequency of $6.835003\,$GHz. In the final $20\,\mu$s, we apply a clean $\pi$ pulse resonant to the $\ket{5S_{1/2},F=1}\rightarrow\ket{5P_{1/2},F=2}$ transition to remove the atoms remaining at $F=1$ after the microwave transition.  With those efforts, we successfully extend the coherence time of each memory mode to $200\,\mu$s level, which is sufficient for our next experiments.

We also characterize the crosstalk between different angular modes within a single memory cell. Here we choose the central memory cell ($103,103$) as the target cell, focusing both write beam and read beam onto this target cell with AOD addressing. We collect the coincidence counts between four signal modes $(i=\{A,B,C,D\})$ and four idler modes $(j=\{A,B,C,D\})$. As shown in Fig. S1c, the obtained results demonstrate that the cross correlation between the signal mode $i$ and the idler mode $j$ is close to 1 when $i\neq j$, indicating that the crosstalk between different angular modes is effectively negligible.

\begin{figure}
  \centering
  \includegraphics[width=15cm]{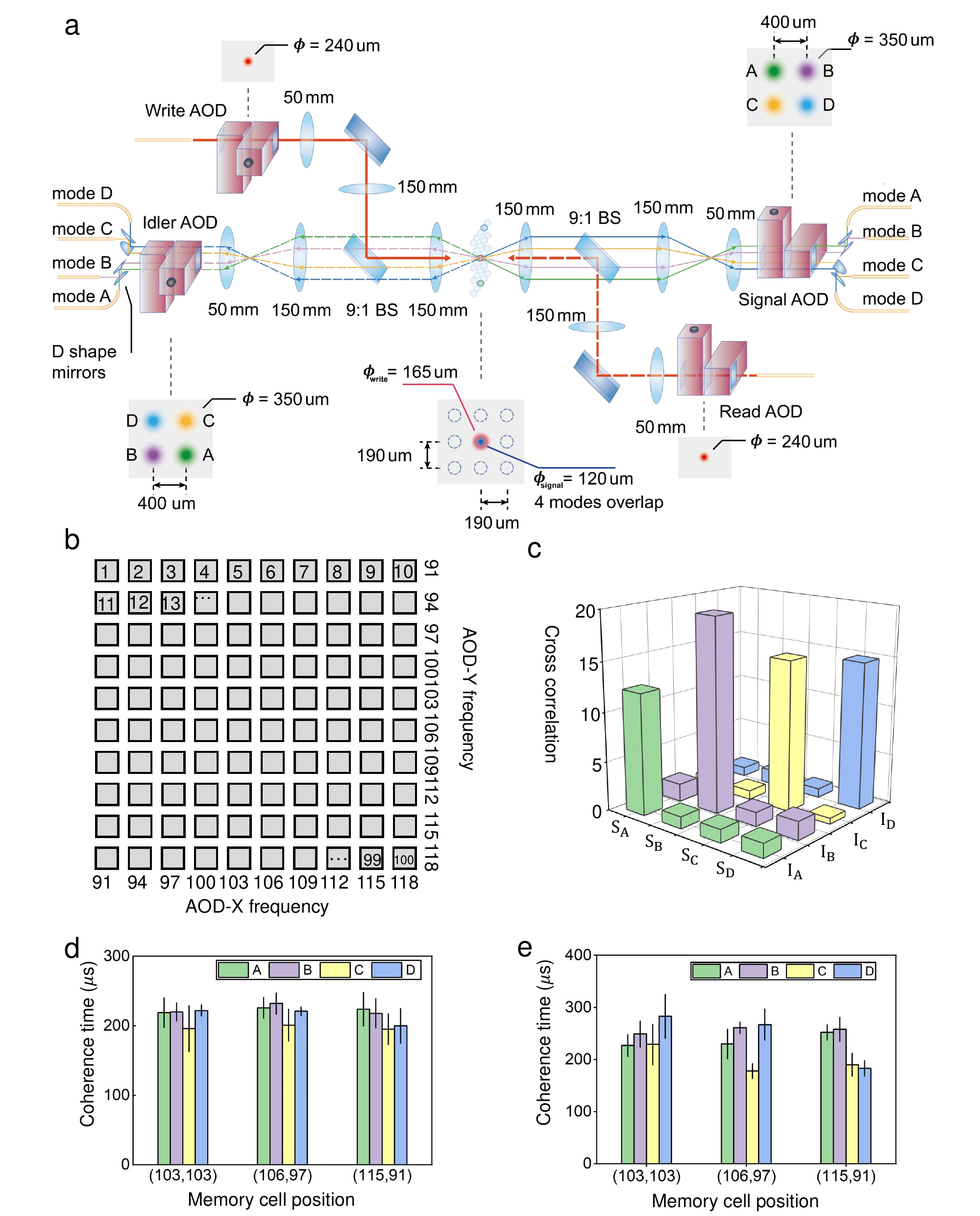}
  \caption{The hybrid-multiplexed quantum memory and performance. \textbf{a}, Detailed experiment setup for the spatially-multiplexed quantum memory array combined with angular degree of freedom. \textbf{b}, This figure shows all the individually addressable memory cells in this work. The AOD addressing frequency for each memory cell is also demonstrated. \textbf{c}, The crosstalk between distinct angular modes in a single memory cell. Here the excitation probability is $0.2\%$ and the storage time is $20\,\mu$s.  \textbf{d}, Here we present the coherence time of 12 modes, which belong to three representative memory cells in Fig. 1d. The coherence time is fitted by cross correlation decay when varying storage time. \textbf{e}, Same as d, but the difference is that the coherence time is fitted by retrieval efficiency decay. }
\end{figure}

\begin{figure}
  \centering
  \includegraphics[width=15cm]{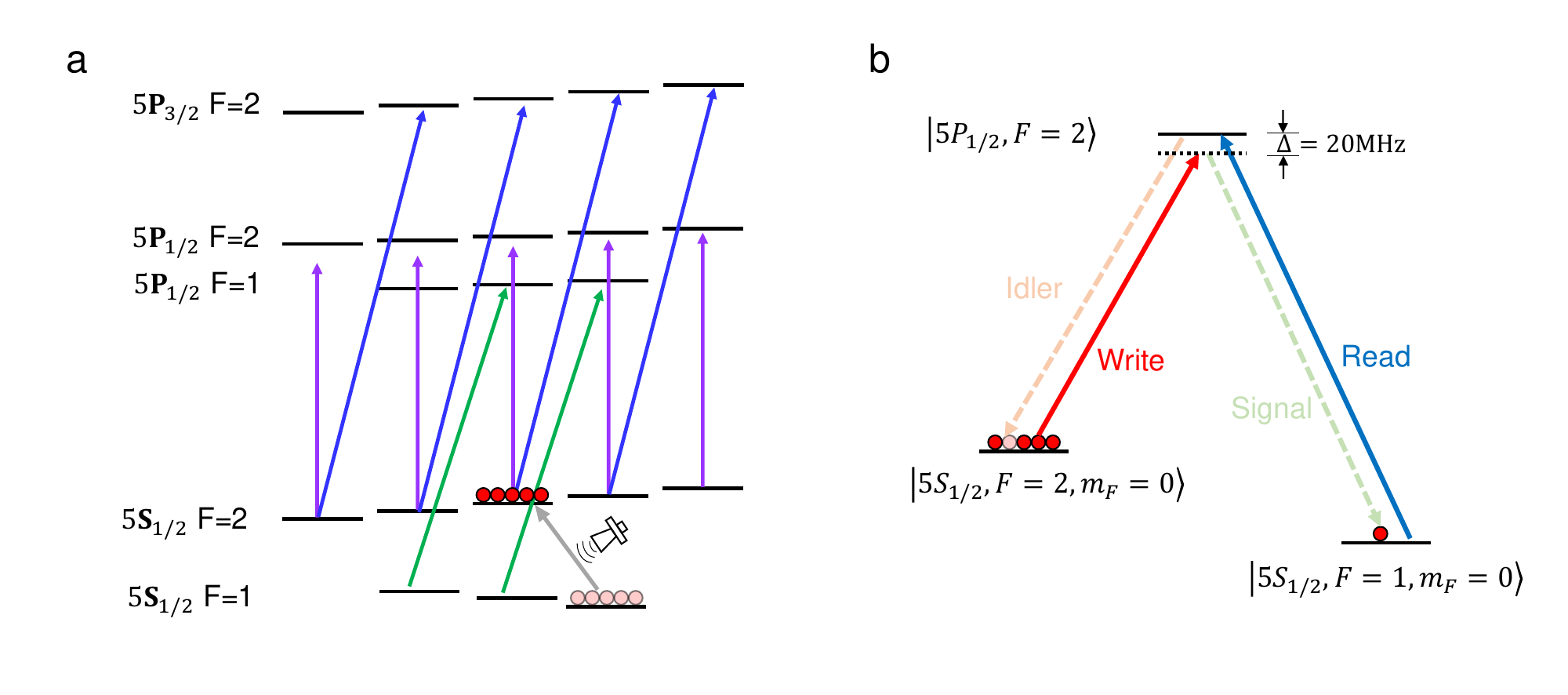}
  \caption{\textbf{a}, The energy levels for the optical pumping process. \textbf{b}, The energy levels for the write process and read process. }
\end{figure}

\section{Implementation of the control logic}

Now we describe the control logic employed to run the delivery of atom-photon quantum correlation over a 12 km fiber in a real-world application. To simulate this scenario, we utilize two optical fibers of equal length, as depicted in Fig. 1b. One is to transmit the converted $1546\,$nm signal photons to the SNSPD, while the other is used for transmitting the classic TTL signal back. This TTL signal serves as an heralding, which informs the memory whether the SNSPD successfully detects the incoming photon.

The experiment sequence begins with a $20\,$ms Magneto-Optical Trap (MOT) loading and a $2\,$ms molasses cooling, followed by $10$ rounds of optical pumping, excitation, and heralding cycles. As illustrated in Fig. S3, the arbitrary waveform generator (AWG) receives an initial trigger signal and subsequently outputs a series of pre-stored waveforms to drive write AOM, write AOD, signal AOD and signal AOMs. During the excitation stage (lasting 120 $\mu$s),  the write beam sequentially visits all memory cells, while simultaneously 280 signal time-bin modes are directed into the telecom fiber. After a round-travel time of $2L/c=120\,\mu$s, the first TTL signal returns, which is received by the Field-Programmable Gate Array (FPGA). During the heralding stage ($120\,\mu$s), the FPGA identifies the excited mode (both the cell and angle) based on the arrival time of the TTL pulse. Upon successful receiving of a TTL pulse, the corresponding index of the excited memory mode is mapped to switch signal to activate the Electro-Optic Modulators (EOMs) and an RF pulse with the corresponding frequency to drive the Read AOM, Read AOD, and Idler AOD. The entire feedforward process is carried out using a home-made FPGA, Digital-to-Analog Converters (DACs), and Voltage-Controlled Oscillators (VCOs). The two different read styles (corresponding to variable storage time and fixed storage time) can be easily switched by programming FPGA.

\begin{figure}
  \centering
  \includegraphics[width=16cm]{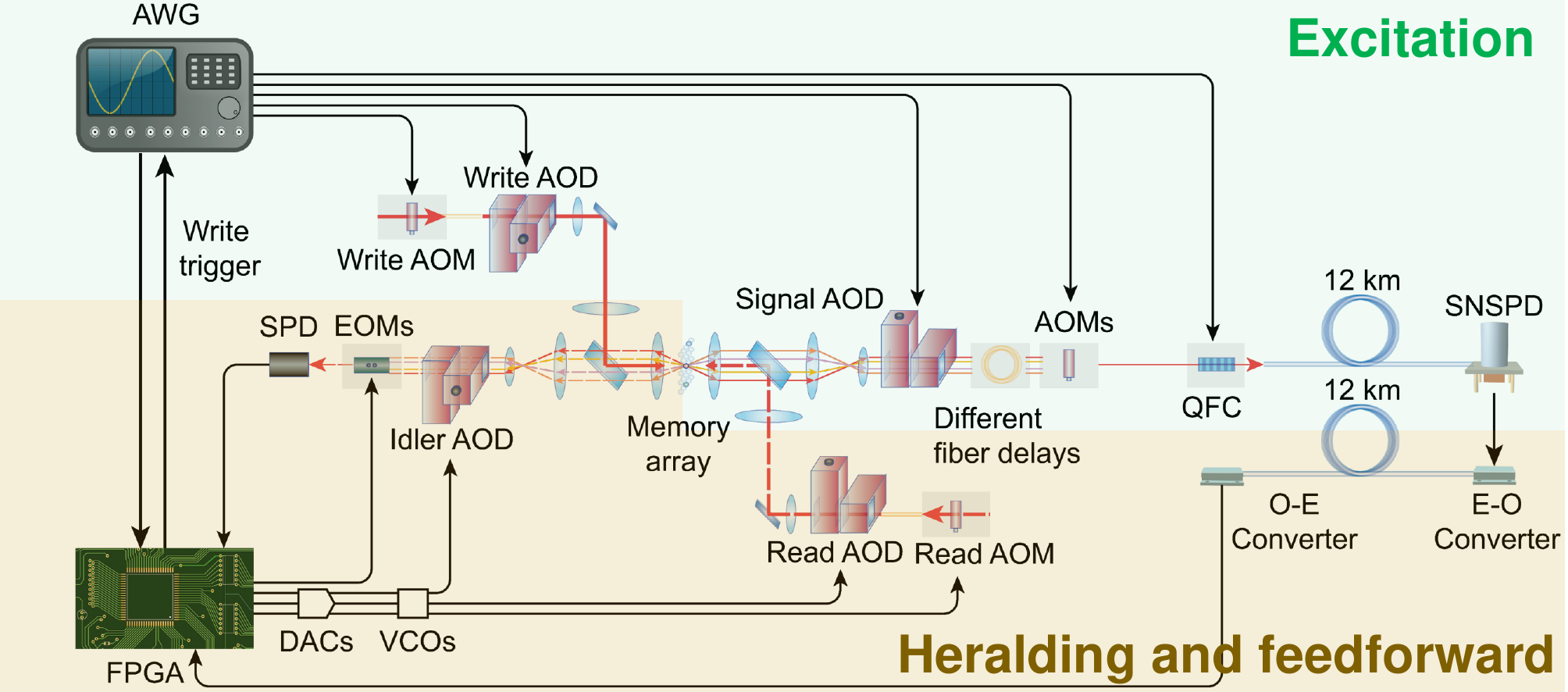}
  \caption{The upper part (light green background) includes all the optical and electronic components working during the excitation stage. The lower part (light yellow background) includes all the optical and electronic components working during the heralding and feedforward stage. Once the MOT loading is finished, the sequential excitation is started.}
\end{figure}

\section{  Different mode numbers in the 12 km fiber}

Here we discuss the experimental sequences for changing the mode numbers in the $12\,$km fiber. As shown in Fig. S4a, when only one memory cell is excited, the total heralding window  (the time span during which heralding signals are expected) contains four photonic time bins. Due to the round-trip travel in two $12\,$km fibers, there exists a constant time difference between the excitation and the subsequent TTL heralding. If the number of modes is changed, the excited memory cells extend from the central region towards the exterior region of the memory array. The reason behind our selection of the lower-left to upper-right regions is for the minimal frequency variation after deflected by the signal AODs, which can cause efficiency variation in the transmission of a narrow etalon (FWHM$\sim100\,$MHz) used for filtering out the pumping laser during the wavelength conversion.

The average storage time varies between the two read styles. The first read-out style (Fig.~3) is to immediately read the corresponding memory cell upon receiving the heralding signal. Consequently, the storage time for all the memory cells is the same. Taking into account the access time $10\,\mu$s required to write the DAC and drive the AOD in the read-out, the fixed storage time of all the memory cells is $\frac{2L}{c}+10=130\,\mu$s. The second read style (Fig. 4) is to read when all the heralding signals have returned. In this case, the average storage time is dependent on the number of used modes. The minimum storage time is still  $130\,\mu$s. The maximum storage time is  $\frac{2L}{c}+\frac{N}{280}\frac{2L}{c}+10=(130+120\frac{N}{280})\,\mu$s if the number of total used modes is $N$. For example, if $N=40$, the maximum storage time is $147\,\mu$s; if $N=280$, the maximum storage time is $250\,\mu$s. In summary, the first read style ensures a fixed storage time for all memory cells, while the second read style results in a variable storage time based on the number of modes used.

\begin{figure}
  \centering
  \includegraphics[width=14cm]{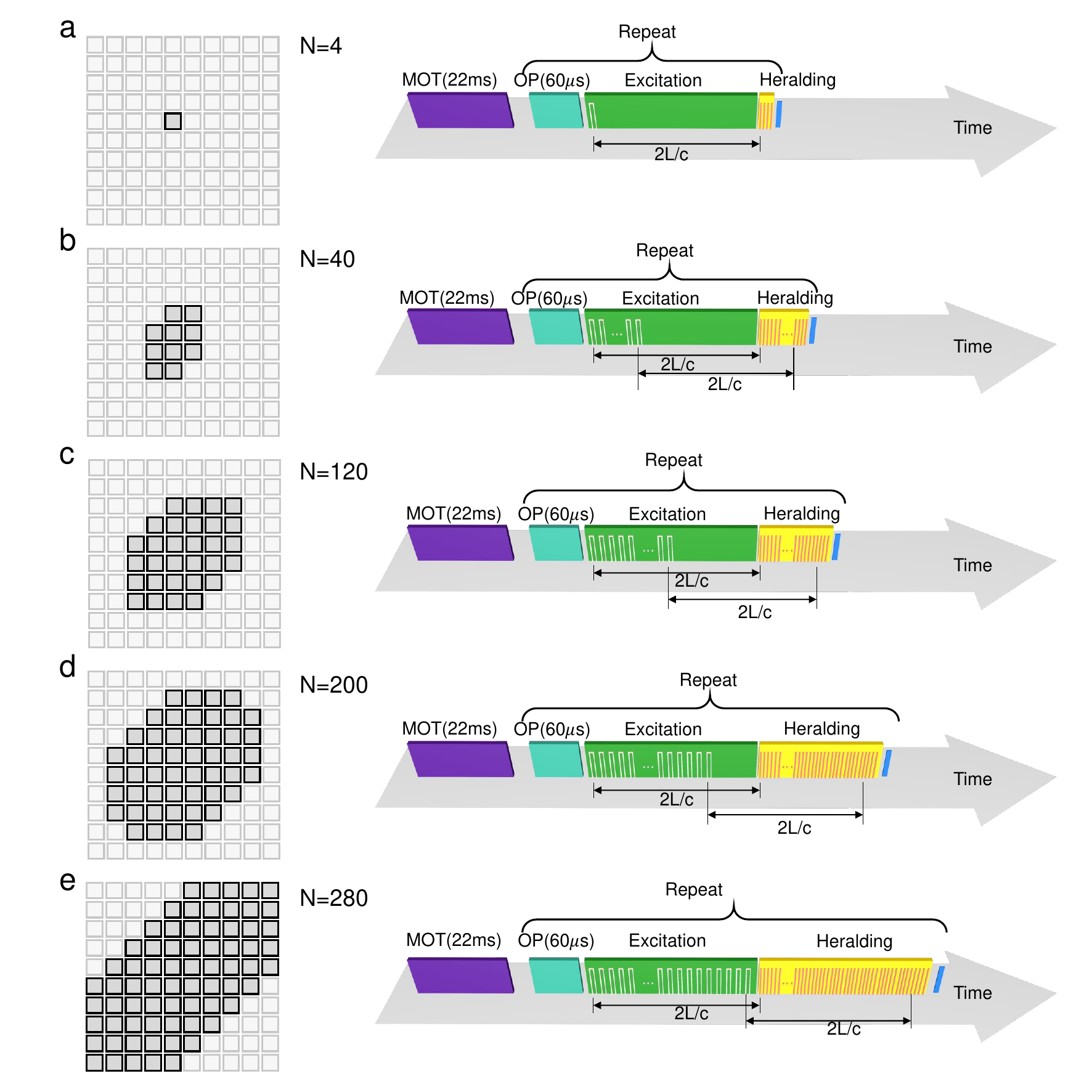}
  \caption{\textbf{a}-\textbf{e}, The selected memory cells and the experimental sequences when changing the memory mode numbers in the 12km fibers. The length of heralding stage varies with different mode number used.}
\end{figure}

\section{ Excitation of 400 modes locally}

As a preliminary experiment, we demonstrate the sequential excitation of all memory cells in a $10\times 10$ array without involving telecom conversion or the transmission through a long fiber. In the absence of the long fiber, the constraint of round-trip communication is eliminated. As illustrated in Fig. S5a, the excitation window aligns with the detection window on the timeline. Given a switching time of $1.7\,\mu$s between different memory cells, the total scanning time for the $10\times 10$ array amounts to $170\,\mu$s, which remains well below the coherence time. Consequently, we excite the entire $10\times 10$ array, thereby generating $400$ pairs of atom-photon correlations locally. Although there isn't a series of photonic time-bin pulses traveling through a long fiber, we are still able to record the arrival times of the $400$ signal photons using the Single Photon Detector (SPD). The histogram in Fig. S5b illustrates the distribution of the $400$ time-bin modes. The excited memory mode is read out at the end of the $170\,\mu$s. By setting the success probability of each mode to $0.17\%$, the average cross-correlation across the $400$ modes reaches $14.8\pm 1.7$.

This preliminary experiment serves as a first step to test the multiplexed protocol for further experiment with longer fiber.

\begin{figure}
  \centering
  \includegraphics[width=14cm]{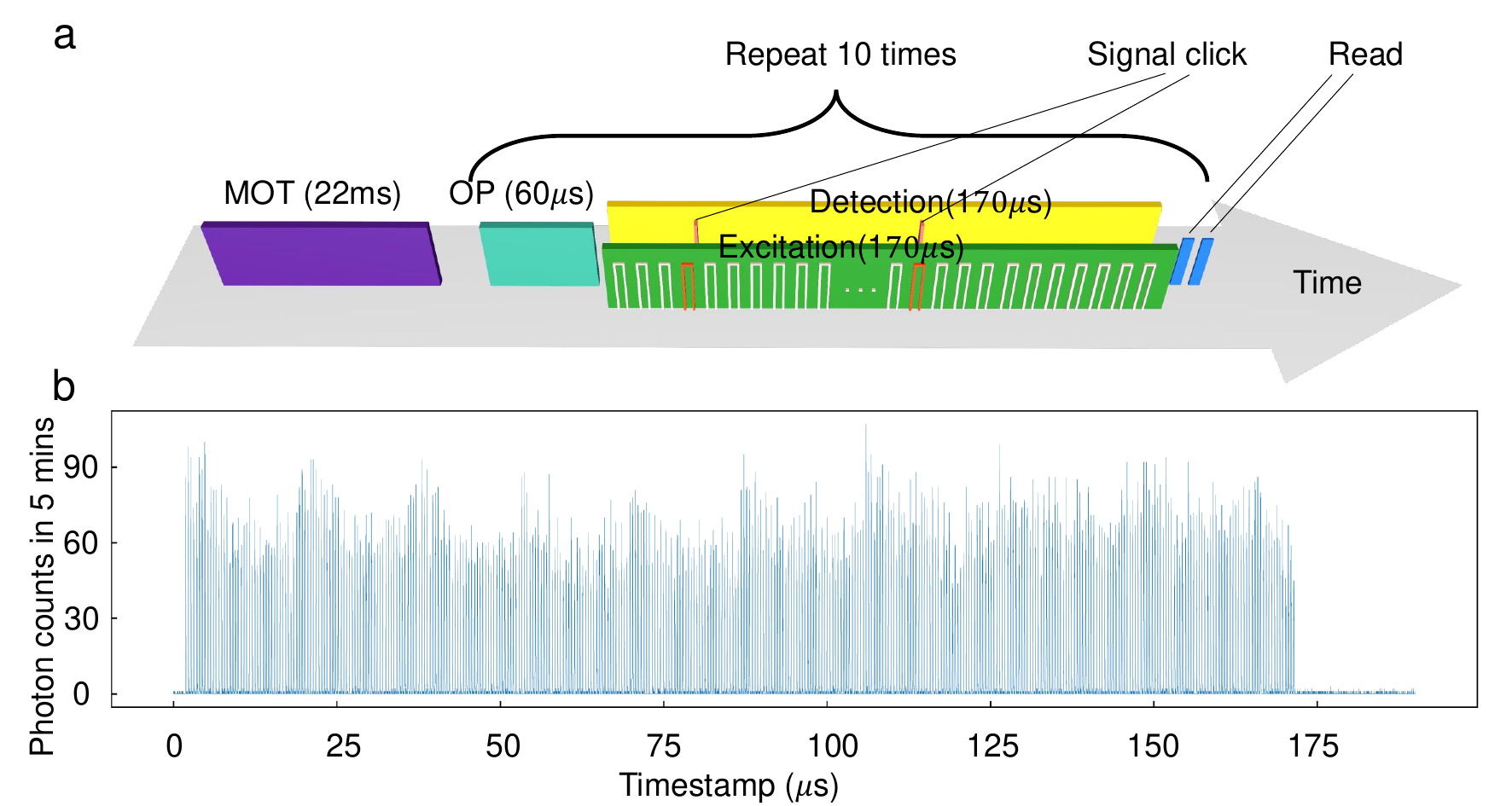}
  \caption{\textbf{a}, The experimental sequence for excitation of 400 modes locally. \textbf{b}, The histogram of the 400 time bin modes recorded by the single photon detector.  }
\end{figure}

\section{Delivery of atom-photon nonclassical correlation with 1km fiber}

The signal photon at $795\,$nm has an attenuation about $4.0\,\text{dB}$/km in the single mode fiber used in our lab, thus can be transmitted with a fiber of $\sim1\,$km.  In our work, we also investigate atom-photon quantum correlations after the signal photons are transmitted through a 1 km fiber without wavelength conversion. The transmission time of the 1 km fiber is approximately 5 $\mu$s, and we use 12 modes in this experiment. Here no frequency conversion is used. Since the scanning time of three cells is shorter than switching time in read-out ($10\,\mu$s), there is no significant difference between the two read styles. Note that here the heralding TTL is directly sent to the FPGA and does not experience the second transmission in the $1\,$km fiber, which is different from the stringent protocol demonstrated in main text. We identified the excited memory mode and retrieved the corresponding idler photon at the end of the detection window. In Fig. S6c, we present the signal photons, idler photons, and coincidence counts for the 12 modes collected over 624 seconds. The average success probability of the 12 modes is $0.1\%$. The average cross-correlation across the 12 modes is $17.1\pm 1.6$, with an average retrieval efficiency of $2.6\%$.

\begin{figure}
  \centering
  \includegraphics[width=14cm]{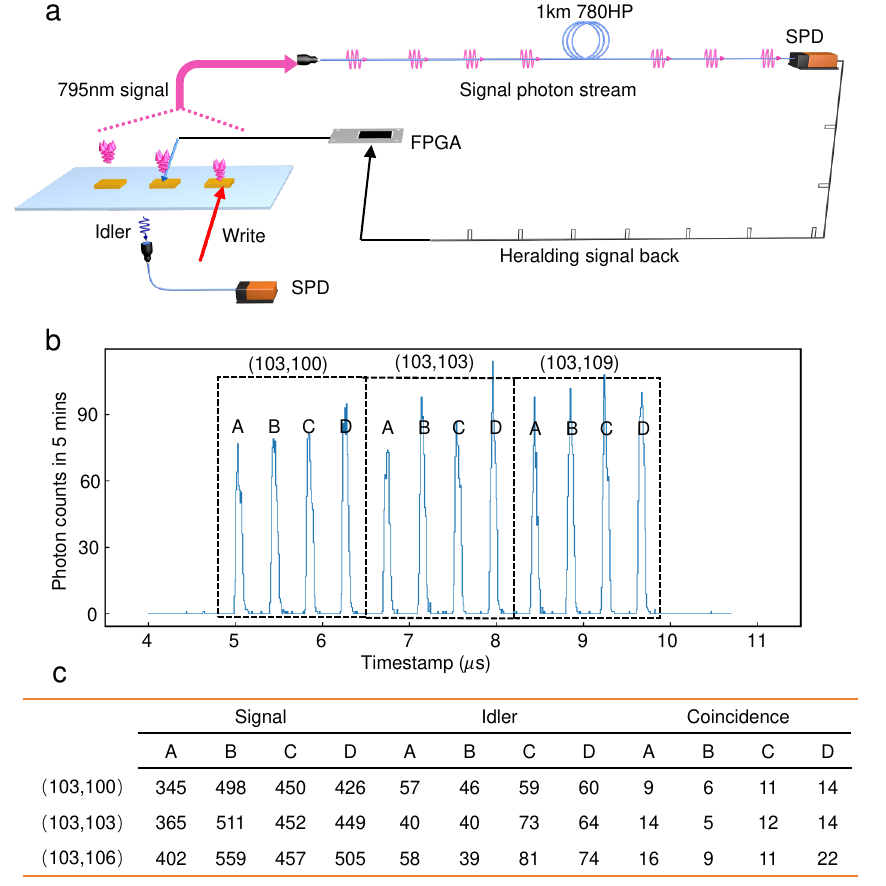}
  \caption{Delivery of atom-photon nonclassical correlation with 1 km fiber. \textbf{a}, Schematic of signal photon transmission over 1km optical fiber. \textbf{b}, The histogram of 795nm signal photon. \textbf{c}, This table shows the signal photon, idler photon and coincidence counts of the 12 modes, collected in 624 s. The average success probability of the 12 modes is $0.1\%$. The memory array (containing 3 cells) is excited 454252 rounds, among which no signal photon is detected for 448848 times.  }\label{1km}
\end{figure}

\section{ Evaluation of the cross correlation for the multimode quantum memory}

Here we outline how to evaluate the average cross correlation over 280 modes, as shown in Fig. 3 and Fig. 4 of the main text. For a DLCZ-type quantum memory, the cross correlation is defines as $g_{s,i}=\frac{p_{s,i}}{p_s p_i}=\frac{p_{(s,i|s)}}{p_i}$, where $p_{(s,i|s)}$ is the conditional probability of detecting signal-idler coincidence given a signal is received. Here we also have $p_i=p_{(s,i|s)}p_s+p_{(\bar{s},i|\bar{s})}p_{\bar{s}}$, where $p_{(\bar{s},i|\bar{s})}$ is the conditional probability of detecting an idler photon during read-out given no signal photon is detected in write, and $p_{\bar{s}}$ denotes no signal photon is detected in write. As $p_s\approx 0.1\%$, $p_{\bar{s}}\approx 1$, and $p_{(s,i/s)}$ is only about $\sim10$ folds larger than $p_{(\bar{s},i|\bar{s})}$, thus $p_i\approx p_{(\bar{s},i|\bar{s})}p_{\bar{s}}$, and $g_{s,i}=\frac{p_{(s,i|s)}}{p_i}\approx\frac{p_{(s,i|s)}}{p_{(\bar{s},i|\bar{s})}p_{\bar{s}}}\approx\frac{p_{(s,i|s)}}{p_{(\bar{s},i|\bar{s})}}$. If we get some signal photon clicks (S) after one round of excitation, we identify and retrieve the excited modes into idler photons to collect the corresponding coincidence counts (C). If no signal photon click is detected after one round of excitation, we record the number of these situations (N) and retrieve one mode randomly selected from all modes into idler photon and collect the idler photon click (I). The average cross correlation across all modes is then estimated to be $g_{s,i}=\frac{p_{(s,i|s)}}{p_{(\bar{s},i|\bar{s})}}=\frac{C/S}{I/N}=\frac{CN}{SI}$. The standard deviation is calculated assuming that all photon clicks follow the Poisson distribution.

\end{document}